\documentclass[iop]{emulateapj}
\usepackage{graphicx}
\usepackage{epstopdf}
\usepackage{color}

\newcommand{\kepler}{{\em Kepler}}
\newcommand{\msun}{$\rm M_{\sun}$}
\newcommand{\rsun}{$\rm R_{\sun}$}
\newcommand{\rearth}{${\rm R_{\earth}} $}
\newcommand{\mearth}{$\rm M_{\earth}$}

\definecolor{orange}{rgb}{.8,0.4,0}
\definecolor{editcolor}{rgb}{0.6,0.0,0}

\newcommand{\tempexpection}{0.36}
\newcommand{\tempexpectionerror}{0.08}

\newcommand{\triggertempexpection}{0.89}
\newcommand{\triggertempexpectionerror}{0.13}

\newcommand{\periodexpection}{0.052}
\newcommand{\periodexpectionerror}{0.015}
\newcommand{\periodexpectionfouryears}{0.21}
\newcommand{\periodpoissonfindingatleast}{19}

\begin{document}

\shorttitle{Planet Occurrence for MEarth's M Dwarfs}
\shortauthors{Berta et al.}
\title{Constraints on Planet Occurrence around Nearby Mid-to-Late M Dwarfs \\from the MEarth Project}
\author{Zachory~K.~Berta\altaffilmark{1}, Jonathan~Irwin, David~Charbonneau}
\affil{Harvard-Smithsonian Center for Astrophysics, 60 Garden St., Cambridge, MA 02138, USA}
\email{zberta@cfa.harvard.edu}

\begin{abstract}
The MEarth Project is a ground-based photometric survey to find planets transiting the closest and smallest main-sequence stars. In its first four years, MEarth discovered one transiting exoplanet, the 2.7\rearth\ planet GJ1214b. Here, we answer an outstanding question: in light of the bounty of small planets transiting small stars uncovered by the \kepler\ mission, should MEarth have found more than just one planet so far? We estimate MEarth's ensemble sensitivity to exoplanets by performing end-to-end simulations of $1.25\times10^6$ observations of 988 nearby mid-to-late M dwarfs, gathered by MEarth between October 2008 and June 2012. For $2-4$\rearth\ planets, we compare this sensitivity to results from \kepler\ and find that MEarth should have found planets at a rate of $0.05 - 0.36$ planets/year in its first four years. {As part of this analysis, we provide new analytic fits to the \kepler\ early M dwarf planet occurrence distribution.} When extrapolating between \kepler's early M dwarfs and MEarth's mid-to-late M dwarfs, we find that assuming the planet occurrence distribution stays fixed with respect to planetary equilibrium temperature provides a good match to our detection of a planet with GJ1214b's observed properties. For larger planets, we find that the warm ($600-700$K), Neptune-sized (4\rearth) exoplanets that transit early M dwarfs like Gl436 and GJ3470 occur at a rate of $<0.15$/star (at 95\% confidence) around MEarth's {later M dwarf} targets. We describe a strategy with which MEarth can increase its expected planet yield by $2.5\times$ without new telescopes, by shifting its sensitivity toward the smaller and cooler exoplanets that \kepler\ has demonstrated to be abundant.  
\end{abstract}

\maketitle

\section{Introduction}\label{s:introduction}

{An overarching goal of exoplanetary science is to understand the physical processes that shape exoplanets smaller than Neptune. Empirical measurements of the masses, radii, and atmospheres of such planets can serve as crucial inputs to developing this understanding. These measurements are most feasible for transiting planetary systems that exhibit favorable planet-to-star mass, radius, and temperature ratios (for amplifying signal strengths) and that are bright (for suppressing photon noise). Planets can satisfy these criteria by transiting nearby small stars: the local M dwarfs. Observations from the \kepler\ mission indicate that early M dwarf stars host, on average, $0.90^{+0.04}_{-0.03}$ planets per star, for planetary radii between $0.5-4$\rearth\ and orbital periods shorter than 50 days \citep[][]{dressing.2013.orspass}. This finding that small planets are so abundant around \kepler's distant M dwarfs bodes well for our prospects of finding such planets transiting {\em nearby} M dwarfs, the close exemplars for which detailed characterization studies will be most rewarding.}

The MEarth Project is an ongoing ground-based survey to identify planets transiting nearby mid-to-late M dwarfs \citep{nutzman.2008.dcgtshpod, berta.2012.tdmsndbcsd}. With MEarth, we discovered the 2.7\rearth, 6.2\mearth\ sub-Neptune GJ1214b \citep[][]{charbonneau.2009.stnls}. GJ1214b transits a very small (0.2\rsun) M dwarf not far from the Sun \citep[14.5 pc away;][]{anglada-escude.2013.1rtpsposbps1}, enabling characterization studies to probe the planet's atmosphere. Transmission spectroscopy from the ground \citep{bean.2010.gtsse1,bean.2011.ontssgfema,croll.2011.btss1smmwa,crossfield.2011.hdnts1,de-mooij.2012.ontos1wm,murgas.2012.nbhps1wgtf,narita.2013.isjstp1, teske.2013.oote1} and from space \citep{desert.2011.oemasg,berta.2012.ftssgfwfchst,fraine.2013.stsgia} is starting to provide boundary conditions  for possible compositions of GJ1214b's gaseous outer envelope, constraints that may prove useful in  our interpretation of the large population of sub-Neptunes unveiled both by Doppler surveys \citep{howard.2010.omdcsnj} and by \kepler\ \citep{batalha.2013.pcokiafmd, fressin.2013.fprkop}

Since the start of the MEarth survey in 2008, GJ1214b is the only planet we have found. If \kepler\ indicates planets are so common around M dwarfs, why has MEarth found only one?  The shape of the occurrence distribution of the \kepler\ M dwarf planets plays strongly into the answer. 
\citet{dressing.2013.orspass} found that while small planets are common around M dwarfs, a planetary radius of 2\rearth\ marks the start of a precipitous decline in the rate of planet occurrence toward larger planets, particularly at short orbital periods. Yet MEarth was specifically designed \citep{nutzman.2008.dcgtshpod} to be sensitive to planets larger than 2\rearth. This mismatch between MEarth's achieved sensitivity and \kepler's implied planet occurrence qualitatively explains MEarth's lack of further detections. In this work, we simulate MEarth's ensemble sensitivity during its first four years, in order to make this statement more quantitative.

Our aim is to provide perspective on the challenges and opportunities for finding planets transiting nearby, very small stars. Due to observational biases, \kepler's M dwarf planet hosts are typically very distant ($100-300$ pc) and are dominated by earlier, larger ($0.3-0.6$\rsun) spectral types. The \kepler-42 trio of sub-Earths transiting a 0.17\rsun\ twin of Barnard's star 40 pc away is a notable but rather rare exception \citep{muirhead.2012.cckisswlpmtsp}. By and large, \kepler\ will not find the Earths, super-Earths, and sub-Neptunes that transit very nearby, very small, mid-to-late M dwarfs, those planets whose atmospheres could be studied in the near future. {Finding these planets requires an all-sky approach, and \kepler's 100 square degree field of view is only 0.3\% of the entire sky.} Focused searches will be required to find these planets, either targeted photometric surveys like MEarth or APACHE \citep{giacobbe.2012.ptspacsfwiaps}, or radial velocity surveys followed by transit monitoring \citep{gillon.2007.dtnn, bonfils.2012.utnd3dwhvctwtp}. As an all-sky survey, the recently accepted Transiting Exoplanet Survey Satellite \citep[TESS;][]{ricker.2010.tesst} mission will make significant progress toward finding planets whose atmospheres can be studied \citep{deming.2009.dctseuatsfjwst}, but is not yet available. The lessons we learn from MEarth and \kepler\ can guide these efforts in the years to come.

In addition to finding individual objects to use as observational laboratories, understanding the statistical properties of the planet population around M dwarfs is important in its own right. Main sequence hydrogen-burning stars with spectral types later than M0 (roughly 0.6\msun) outnumber G dwarfs like the Sun a dozen to one in the solar neighborhood \citep{henry.2006.snxprfcpmrps}\footnote{\url{See also http://www.recons.org/census.posted.htm}}. {Estimates from SDSS indicate that the Galaxy's stellar mass function peaks at 0.25\msun\ \citep{covey.2008.lmflsgdicr,bochanski.2010.lmflsgdif}, corresponding roughly to M4 in spectral type. The \kepler\ M dwarfs probe planet statistics down to this peak, but not much below it. In the SDSS mass function, 40\% of stars fall below 0.25\msun. If we want a complete census of exoplanets in the Milky Way, we need to understand how planets populate these smallest of stars. }

{Observations of how planet populations change with host star mass inform theories of planet formation and evolution. For example, \citet{laughlin.2004.campjpod} argued that, in the core accretion paradigm, M dwarfs' less massive protoplanetary disks \citep[see][]{andrews.2005.cddtsp} would inefficiently form Jupiter-mass planets. This hypothesis was later verified observationally as a deficit of Jupiters detected in radial velocity surveys \citep{johnson.2007.padrcbesm}. By spanning a wide range of stellar masses, we can hope to provide a long lever arm to probe how the birth, growth, and survival of planets depends on various environmental factors \citep[see][]{raymond.2007.dphpfals,barnes.2008.teph,montgomery.2009.fdempams,ogihara.2009.nspaads}. }

In this work, we use MEarth's single planet detection and lack of additional detections to explore the statistics of planets at the bottom of the main sequence. The paper is organized as follows. In \S\ref{s:observations}, we describe four years of observations gathered by the MEarth Project. In \S\ref{s:method}, we outline our numerical simulations of MEarth's sensitivity during this time. In \S\ref{s:results}, we discuss the results and compare them directly to the M dwarf planet statistics from \kepler. In \S\ref{s:future}, we demonstrate how to harness these findings to improve MEarth's planet yield in the years to come. We conclude in \S\ref{s:conclusions}.

\section{Observations}\label{s:observations}
MEarth employs eight robotic 40-cm telescopes at the Fred Lawrence Whipple Observatory atop Mt. Hopkins in Arizona. { The cameras are sensitive to 715--1000nm wavelengths, with the bandpass shaped by a fixed Schott RG715 filter\footnote{A more narrow filter with an interference cutoff (715-895m) was used in the 2010--2011 season; see \citet{berta.2012.tdmsndbcsd}.} and the quantum efficiency of our back-illuminated e2v CCD42-40 detectors}. The light curves analyzed in this paper were gathered between 2 October 2008 and 23 June 2012. Except for the annual monsoon season typically spanning July to September and occasional instrumental failures, MEarth observed every clear night during these four years. \citet{nutzman.2008.dcgtshpod} outline MEarth's design strategy, and descriptions of the light curve processing and properties can be found in \citet{irwin.2011.amefcsrpfmfmts}, \citet{berta.2012.tdmsndbcsd}, {and on the MEarth website\footnote{\url{http://www.cfa.harvard.edu/MEarth/}}.}

MEarth differs from wide-field transit searches. Targeting the brightest mid-to-late M dwarfs spread across the sky, MEarth observes each star one-by-one in a pointed fashion. As such, MEarth can tailor exposure times to individual targets; we observe each target long enough in each visit that its expected RMS from photon and scintillation noise is 1/3 the transit depth of a 2\rearth\ planet. {If necessary (to avoid saturation), we gather multiple exposures within each visit to meet this goal.} For the estimated $0.1-0.35$\rsun\ radii of MEarth targets \citep{nutzman.2008.dcgtshpod}, these correspond to photometric precisions of $1.0-0.2\%$ {per visit}. Thanks to a comparatively small pixel scale (0.76"/pixel) and the large proper motion on which our targets were selected \citep[$\mu > 0.15"/{\rm year}$; from][]{lepine.2005.cnswapmlt0lc}, blended background eclipsing binaries are not a significant source of false positives \citep[see][for contrast]{latham.2009.dtpeebhfg, triaud.2013.wgtcd}.

\begin{figure}[tbp]
   \centering
   \includegraphics[width=\columnwidth]{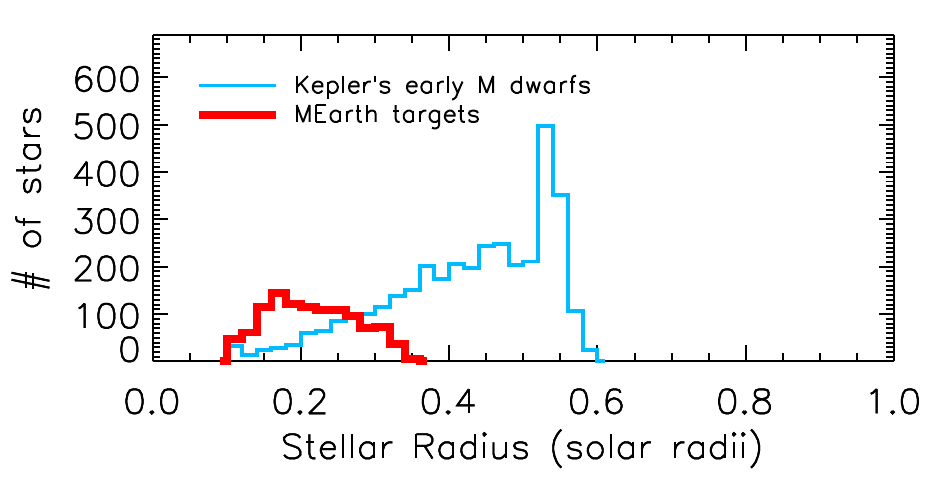} 
   \caption{{Comparison of MEarth targets ({\em red}) with the \kepler\ early M dwarf sample ({\em blue}). Stellar radius estimates are taken from \citet{nutzman.2008.dcgtshpod} for MEarth and from \citet{dressing.2013.orspass} for the \kepler\ M dwarfs.} Only MEarth targets analyzed in this work are shown, not the entire  \citet{nutzman.2008.dcgtshpod} sample.}
   \label{f:radii}
\end{figure}

MEarth's cadence is also unique. Transits of these stars have typical durations of $0.5-2$ hours. MEarth observes each star roughly once every {20--30} minutes, reduces the data in realtime, and can trigger immediate high-cadence monitoring of interesting in-progress transit events. By gathering many observations of a transit's egress with this realtime trigger mode, MEarth can confirm the presence of a planet from a single transit. As we discuss in \S\ref{s:future}, this mode can substantially improve MEarth's sensitivity to long period planets. The realtime trigger was operational for most of $2009-2012$, but was under continual development to ramp up its sensitivity during this time. 

Figure \ref{f:radii} shows the estimated radii of the MEarth stars analyzed in this paper. For comparison, the \kepler\ early M dwarf sample analyzed by \citet{dressing.2013.orspass} is also included. The \kepler\ M dwarfs are strongly biased toward earlier spectral types, near the K/M boundary. The radius estimates for the MEarth sample are drawn from \citet{nutzman.2008.dcgtshpod}. They are highly uncertain, but we address this issue in our analysis in \S\ref{s:stellar_errors}.

All M dwarf light curves from MEarth are made publicly available for the community to use. Data are released on a yearly schedule, and can be found online at \url{http://www.cfa.harvard.edu/MEarth/}. {The online release contains extensive notes on the data processing methods, and on the properties of the survey as a whole.} We eagerly encourage those interested in the photometric variability of nearby mid-to-late M dwarfs, or the possible presence of planets around them, to make broad use of {MEarth's public} light curves.

\subsection{Data Quality Cuts}
{For the analyses presented in this work, only M dwarf light curves consisting of more than 100 good visits are included.  For the four seasons since 2008, there were  $[349,306,256,240]$ such light curves. We excluded from the statistical analysis all
light curves where genuine astrophysical eclipse signals were detected
during the course of the survey, as they are dominated by high cadence follow-up observations and do not reflect our blind survey sensitivity. These objects are GJ1214 \citep{charbonneau.2009.stnls}, GJ 3236 \citep{irwin.2009.3bvmebsdmo}, NLTT 41135 \citep{irwin.2010.n4fdbdebtsdmo}, and LSPM J1112+7626 \citep{irwin.2011.ljdmebfmts}.

Most stars were observed for only a single season, but not all. Some M dwarfs (140 in total) were observed in two, three, or four seasons. One season typically dominates the sensitivity to planets in these cases; for simplicity, we include only the most sensitive season-long light curve for each star in the statistical analysis, thus slightly underestimating our true sensitivity.  In total, 988 unique M dwarfs are represented here.

We make strict data quality cuts. We flag and remove any exposures in which thick clouds were present (worse than 0.25 magnitudes of losses), the target's position was offset from its median on the detector by more than $5\sigma$, images show extremely large FWHM ($3\sigma$ above its median) or high ellipticity ($>0.25$),  $>10\%$ of comparison stars show $4\sigma$ outliers in their light curves, or $50\%$ of the other exposures within a 4 hour window were excluded for any other reason. Some exposures, originally identified as contributing to candidate eclipse signals, were found to contain anomalies (inopportune glints or satellite tracks, shutter blade failures, other defects) and were also excluded.  A total of $1.25\times10^6$ exposures distributed over $8.26\times10^5$ independent visits survive these cuts and contribute to the analysis.}

\section{Method}\label{s:method}
We aim to interpret MEarth's one planet detection and many non-detections in the light of the statistical rate of occurrence around these small stars. Doing so requires an estimate of the ensemble sensitivity of the survey to hypothetical planets during the past four seasons. 
\subsection{MEarth's Ensemble Sensitivity}
We define MEarth's ensemble sensitivity, $S(R,P)$, as the number of planets with a particular radius $R$ and orbital period $P$ that MEarth should have found if all stars in our sample had one planet with that radius and period. We calculate the ensemble sensitivity as
\begin{equation}
S(R,P) = \sum_{i=1}^{N_{\star}} \eta_{{\rm tra}, i}(R,P) \times \eta_{{\rm det}, i} (R,P)
\label{e:sensitivity}
\end{equation}
where the functions $\eta_{{\rm tra}, i}(R,P)$ and $\eta_{{\rm det}, i} (R,P)$ are the per-star transit probability and the per-star detection efficiency, respectively. Our targets span a factor of $3$ in stellar radius, and each star has its own observational coverage and light curve properties. As such, we calculate each of these functions individually on a star-by-star basis. {The sum is over the $N_\star=988$ stars in the sample.}

In addition to $S(R,P)$, we also calculate our sensitivity as a function of the equilibrium temperature $T$ that a planet on a circular orbit would have, assuming zero albedo and efficient heat redistribution. We refer to this sensitivity as $S(R,T)$ and will use the two ``period-like'' parameters $P$ and $T$ somewhat interchangeably hereafter. 
 {In our definition, $T$ is identically related to the bolometric flux the planet receives from the star and has a one-to-one correspondence with $P$, given each star's estimated mass and luminosity\footnote{We quote temperatures for a Bond albedo of $A=0$ simply for ease of scaling. The relation that $T\propto(1-A)^{1/4}$ can be used to translate to a chosen planetary albedo.}. We use $T$ simply to describe a planet's rough energy budget; the physical conditions throughout the planet will depend on many more factors than can be encapsulated by a single parameter. For example, the shape of the star's input spectral energy distribution can have a strong influence on a planet's surface level atmospheric properties and (at temperatures cooler than those primarily probed in this paper) habitability \citep[see][]{kopparapu.2013.hzamse,rugheimer.2013.sfepas}; our $T$ should not be interpreted more deeply than as a useful shorthand for the amount of energy impinging the top of a planet's atmosphere. }

\subsection{Transit Probabilities = $\eta_{{\rm tra}, i}(R,P)$}
The first factor in Equation \ref{e:sensitivity}, $\eta_{{\rm tra}, i}(R,P)$, is the probability that a planet is geometrically aligned such that transits occur in the system. We take $\eta_{\rm tra}=R_\star/a$, where $R_\star$ is the stellar radius and $a$ is the semi-major axis; this is the probability that a planet in a circular orbit would have an impact parameter $b<1$. From Kepler's third law {and assuming the planet's mass is negligible}, $R_\star/a = [3\pi/(GP^2\rho_\star)]^{1/3}$ where $G$ is Newton's constant. We estimate the mean stellar density $\rho_{\star}$ from archival observations of each target \citep[see][]{nutzman.2008.dcgtshpod}. We discuss the impact of uncertainties in $\rho_\star$ on our $S(R,P)$ in \S\ref{s:stellar_errors}.

Non-zero eccentricity can increase (or decrease) transit probabilities for planets with Earth-ward periastrons (or apoastrons), but the net effect for blind transit surveys is small on average. For a reasonable underlying distribution of eccentricities, assuming circular orbits underestimates the total sensitivity of a photon-limited survey by only $4\%$ and of a completely red-noise dominated survey by $25\%$ \citep{burke.2008.ioedtep}. With our sparse sampling, our ability to correct for known systematic effects, and the short $1-2$ hr transit durations we are targeting, MEarth's sensitivity is influenced but not dominated by red noise \citep{berta.2012.tdmsndbcsd}. As such, we expect the eccentricity bias to be less than about $10\%$ and ignore it.

\begin{figure*}[htbp] 
   \centering
   \includegraphics[width=\textwidth]{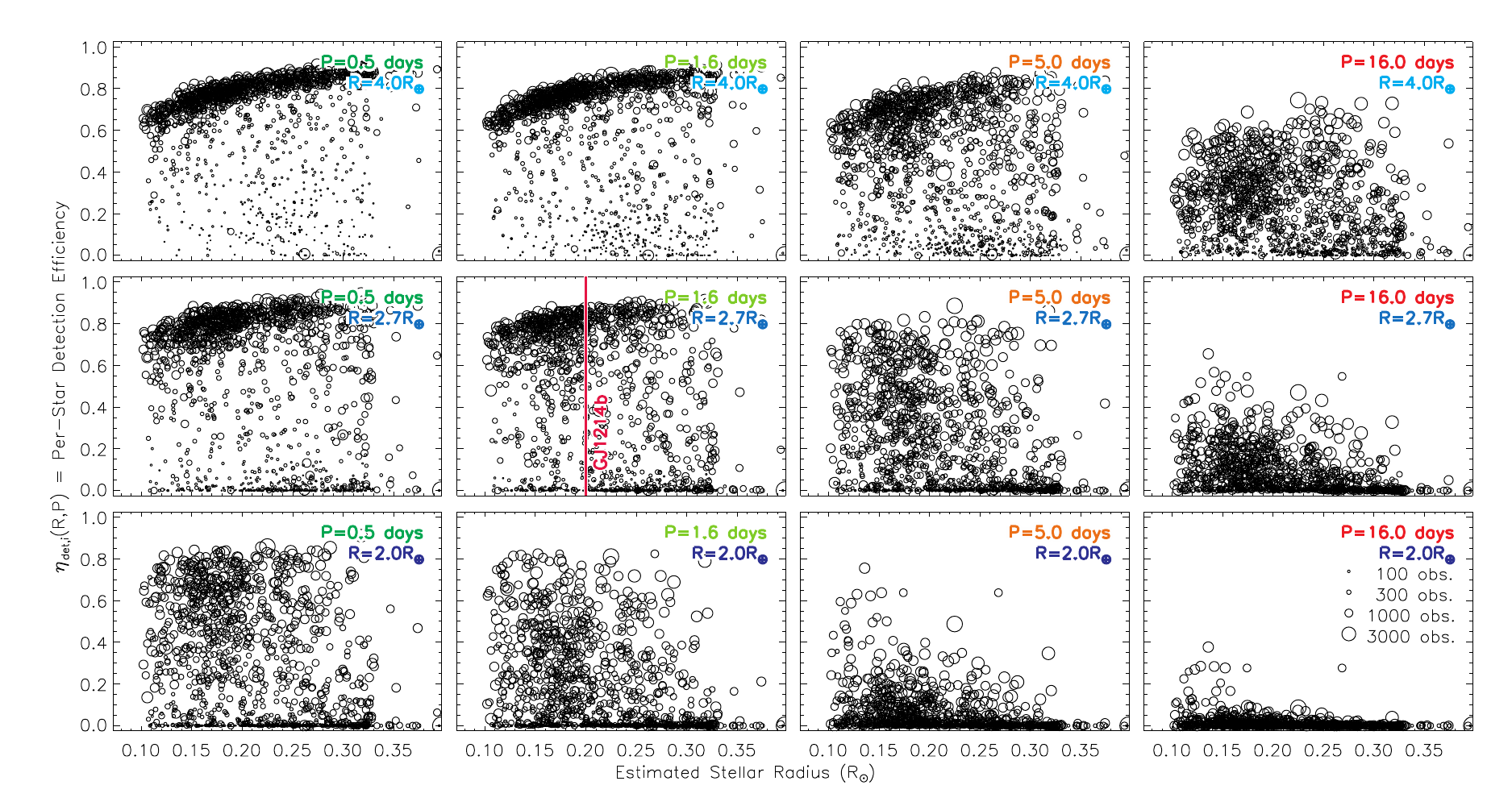} 
   \caption{Results from the simulation described in \S\ref{s:detectionefficiency}, showing the per-star discovery efficiency $\eta_{\rm det, i}$ in a phase-folded search, for planets of decreasing radii ({\em rows}, [4.0, 2.7, 2.0]\rearth\ top to bottom) and increasing periods ({\em columns}, [0.5, 1.6, 5.0, 16] days left to right). Each star is represented by a single circle appearing in each of the 12 panels; the area of each circle is proportional to the number of observations MEarth  gathered of that star. Transits become more difficult to detect for smaller planets and longer periods. }
   \label{f:sample}
\end{figure*}

 \subsection{Transit Detection Efficiencies = $\eta_{{\rm det}, i} (R,P)$}\label{s:detectionefficiency}
The second factor,  $\eta_{{\rm det}, i} (R,P)$, is the probability that a planet aligned to transit would exhibit one or more such transits while MEarth was observing and that the signal would be strong enough cross an appropriate detection threshold. For idealized surveys,  $\eta_{{\rm det}, i}$ can be calculated analytically \citep{pepper.2003.uasfpt,gaudi.2005.pdcegp}, given a typical photometric precision and number of observations. Kepler's detection efficiency appears to match these predictions quite well \citep{youdin.2011.ecgmak, howard.2012.pow0ssfk}. Without the benefit of \kepler's uniform cadence, coverage, and noise properties, we turn to simulations to estimate  $\eta_{{\rm det}, i}$ for MEarth.

Into each MEarth light curve, we inject simulated, limb-darkened transit signals \citep{mandel.2002.alcpts}. The fake planets are generated on a grid of planetary radii spanning 1.5--4\rearth, and have random periods between 0.25 and 20 days, random epochs, and random impact parameters between $b=0$ and $b=1$ (all drawn from uniform distributions). In \citet{berta.2012.tdmsndbcsd}, we develop a framework for assessing the significance of transit signals in MEarth data. This framework, the Method to Include Starspots and Systematics in the Marginalized Probability of a Lone Eclipse (MISS MarPLE), is designed to provide robust detection statistics for transit signals amidst systematic noise sources arising from instrumental/telluric effects or from stellar variability. We inject the signals into our basic light curves\footnote{By ``basic light curves'', we mean MEarth light curves after the differential photometry correction has been applied. This differential photometry correction is estimated solely from comparison stars, not the target M dwarf itself, so any suppression of transits at this early stage of the processing should be minimal.}, process those light curves with MISS MarPLE, and estimate the signal-to-noise ratio at which the injected signal would be recovered. We iterate this process 50,000 times for each star, to map out the planetary parameter space.

We estimate the transit significance at the known period and epoch of the injected signal; we do not perform a phased period search for each injected signal. Because we skip the computationally costly period search, these simulations know nothing about whether or not the existing MEarth light curve would be enough to recover the correct period for the planet. This is okay; we are interested merely in whether or not a strong enough signal would exist in the data that we would track it down with targeted follow-up observations. 

A crucial component is the decision of a detection significance threshold.  MISS MarPLE both corrects for stellar variability and instrumental systematics and keeps track of the excess uncertainty these corrections introduce to the marginalized transit depth uncertainties ($\sigma_{\rm MarPLE}$) of candidate transit signals. In that work, we found that all period-phased MEarth candidates above a transit depth signal-to-noise of $D/\sigma_{\rm MarPLE} >7.5$ were either catastrophic instrumental problems, obvious stellar variability artifacts, or bonafide eclipsing systems \citep{charbonneau.2009.stnls, irwin.2009.3bvmebsdmo,irwin.2010.n4fdbdebtsdmo,irwin.2011.ljdmebfmts}. GJ1214b was identified in 2009 before the development of the full MISS MarPLE framework, using the Box-fitting Least Squares search algorithm \citep{kovacs.2002.baspt} and pre-cleaning with an iterative median filter \citep{aigrain.2004.ppp} and Trend Filtering Algorithm \citep{kovacs.2005.tfawvs}. Reanalyzing the data available up to the time of GJ1214b's original identification, we find that MISS MarPLE would have recovered the signal with $D/\sigma_{\rm MarPLE} = 8.2$. 

We take $D/\sigma_{\rm MarPLE} >7.5$ as our detection threshold for the following analysis. We emphasize that this is a {\em phased} transit significance, meaning that multiple transits can folded together to build up $D/\sigma = 7.5$. For Neptune-sized planets, a single transit observed at 25 minute cadence can often cross this threshold by itself. We impose no lower limit on the number of transits required for a detection, so such planets would be considered as recovered. For the main results in \S\ref{s:results}, we do not explicitly model detecting planets from single transits with MEarth's realtime trigger mode.  Because the realtime trigger was continually being improved throughout the survey, it is difficult to quantify the extent to which it boosted our sensitivity. In \S\ref{s:results}, we choose to include data that were gathered as part of high-cadence follow-up, but we do not simulate the generation of new triggers. As such, MEarth's actual sensitivity may have been higher than we quote. We explore the effect that the realtime trigger can have on our sensitivity in \S\ref{s:future}.

Note that in this context, ``detected'' at $7.5\sigma$ means that a candidate would have high enough significance that, after a thorough visual vetting, we would begin follow-up observations to confirm the transits and to measure the system's period.  Using a detection threshold lower than $7.5\sigma$ would make MEarth appear more sensitive, but drawing statistical inferences would require that we prove the candidates above that threshold were either real or statistical flukes. We are in the process of vetting the most promising $6.5-7.5\sigma$ candidates, but have not yet determined whether they are bona fide planets or not. For simplicity, we simply adopt the more conservative $7.5\sigma$ threshold. We explored changing the assumed detection threshold down to $7\sigma$ and up to $8\sigma$. We found that the quantitative results in \S\ref{s:results} and \S\ref{s:future} do change, but the qualitative conclusions do not.

Figure \ref{f:sample} visualizes the results of our simulations of the phased-search per-star detection efficiency $\eta_{\rm det, i}$. Each panel corresponds to a different slice of planetary radius and period, and contains one circle for every star in the sample. As expected, MEarth's ability to discover planets in a phased search falls off toward longer periods and smaller radii. The symbol area scales linearly with the number of observations MEarth gathered of that star, demonstrating how $\eta_{\rm det, i}$ grows with the number of measurements gathered, and saturates for large radii and short periods. The simulations have effectively integrated over impact parameters $0<b<1$. Because $\eta_{\rm det, i}$ accounts for grazing eclipses being more difficult to detect, it never reaches unity. The downturn in the upper envelope of  $\eta_{\rm det, i}$ toward the smaller stars for short-period, Neptune-sized planets is  caused by the breakdown of MISS MarPLE's assumption that transits will be box-shaped. The mismatch in shape would be treated as excess nightly noise in the MarPLE framework, and these transits' significance down-weighted. A more flexible trapezoidal assumption for the light curve shape \citep{aigrain.2007.mpsoyoc} would be able to improve sensitivity in this corner of parameter space.

\subsection{Uncertainties in Stellar Parameters and Binarity}\label{s:stellar_errors}
MEarth's actual sensitivity depends on how well we understand the physical masses and radii of the stars in our sample. For this work, we adopt the stellar parameters as derived in \citet{nutzman.2008.dcgtshpod}. These parameter estimates {are not necessarily the best estimates currently possible, and introduce uncertainties into our statistical analysis. In this section, we describe those uncertainties and quantify their overall effects.}

First, the masses and radii are imprecise. Although some stellar parameter estimates were based on literature geometric or spectroscopic parallaxes, many are based on {photographic plate measurements with substantial uncertainties.} High proper motions ensure the sample is free of giants, but the roughly 30\% uncertainties in stellar radii directly translate into comparably large uncertainties in the planetary radius to which a given transit depth corresponds. 

Second, some stars may be contaminated with unresolved binaries. We made small efforts to remove some known binaries from our sample, such as those that \citet{lepine.2005.nsflpcimdgwp} identified as visual binaries resolved on input photographic plates or those indicated as spectroscopic binaries in the literature \citep[e.g.][]{shkolnik.2010.tlsb}. However, these efforts took place concurrently while we were already gathering data and are still far from complete. MEarth definitely observed some contaminating binaries as part of the sample analyzed here. The additional flux from a binary companion dilutes transit depths, as well as perturbing the stellar parameters that we infer from (spectro)photometry. 

These two effects have impacts on both $\eta_{\rm tra, i}$ and $\eta_{\rm det,i}$. To prevent them from biasing our ensemble sensitivity estimate, we simulate their influence in a Monte Carlo fashion and apply a correction to our final estimate of $S(R,P)$. In these simulations, we perturb the assumed mass of each star by a 30\% Gaussian uncertainty combined with a prior so that the star is drawn from the local stellar mass function \citep{bochanski.2010.lmflsgdif}. Including the stellar mass function tends to increase the masses of those stars that were originally inferred to be the least massive, those below the mass function peak. We use the mass-radius-luminosity relations of \citet{boyajian.2012.sdtimm} to generate new radii and luminosities from these masses. We randomly assign binary companions to each star, assuming a 34\% multiplicity rate \citep{fischer.1992.mad, janson.2012.almms} and uniform mass-ratio distribution (treating the original star as the primary). We shift the values of $\eta_{\rm tra, i}$ and $\eta_{\rm det,i}$ according to the simulated system's updated properties. We apply this process to all the stars included in the sample, calculate the updated $S(R,P)$ and $S(R,T)$, and repeat 20 times for the ensemble. {We take the averages of these distributions as our best-guess sensitivity estimates.}

\begin{figure*}[htbp] 
   \centering
   \includegraphics[width=\textwidth]{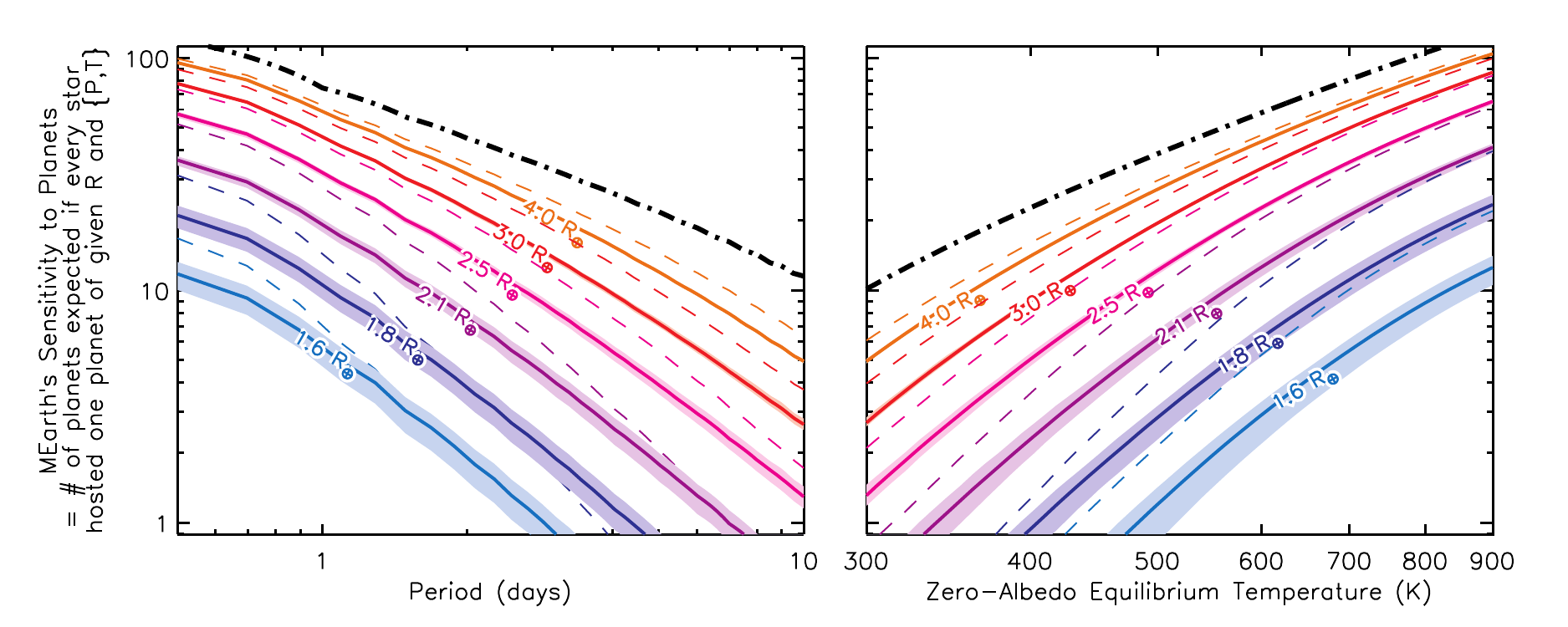} 
   \caption{MEarth's ensemble sensitivity, as a function of planet radius ({\em colors}), orbital period ({\em left panel}), and bolometric incident flux ({\em right panel}), as given by Eq. \ref{e:sensitivity}. These sensitivity curves include both the geometrical probability that a planet would transit and the probability that we would detect transits that do occur. We show MEarth's sensitivity after accounting for uncertainties in the inferred stellar parameters and individual targets' unknown binarity ({\em solid lines}, with the range of values spanned in our perturbation analysis represented by the shaded swaths). For comparison, we show the naive estimate of what the sensitivity would be if we assumed all our targets were single and had perfectly known stellar parameters ({\em dashed lines}). The window function ({\em black dash-dotted line}) shows how sensitive MEarth would be if all that was required to detect a planet was a single in-transit point.}
   \label{f:sensitivity}
\end{figure*}

\section{Results}\label{s:results}

The final sensitivity estimates are shown in Figure \ref{f:sensitivity}, both as a function of orbital period [$S(R,P)$] and as a function of planet's zero-albedo equilibrium temperature [$S(R,T)$]. Because the MEarth stars span a range of effective temperatures, the mapping between these two parameter spaces is not necessarily one-to-one. 

The detectability of planets near or below 2\rearth\ depends strongly on the inferred stellar radius of each star, so the above perturbation analysis introduces substantial uncertainties into the sensitivity for these smaller planets. In contrast, 4\rearth\ planets are often far above the detection threshold, so sensitivity depends weakly on the exact stellar radius. In most cases, the post-perturbation sensitivity decreased relative to the naive estimates. However, for planets smaller than 2\rearth, the sensitivity increased relative to the naive estimate, thanks to some stars stars having radii smaller than originally assumed, so that MEarth's exposure times designed to detect 2\rearth\ planets could actually pick up even smaller planets. 

{Throughout this section, we test a variety of hypotheses against these sensitivity curves. Sometimes, we consider planets of particular periods $P$, other times planets with particular equilibrium temperatures $T$. We explore both parameters because we do not know the answer to the following question: When comparing samples of planets orbiting stars of different masses, what parameters define a planet population? Are two planets orbiting different stars comparable  if they share the same radius? Or should they have the same mass, the same planet-to-star mass ratio, the same ratio to the isolation mass \citep[e.g][]{ida.2008.tdmpfvanls}, or some other shared trait? Are two planets similar if they have the same period, the same semimajor axis, the same $a/R_\star$, the same proximity to the snow line, or something else? Theory points to the importance of various of these parameters in formation, migration, and evolution scenarios \citep[for a small selection of examples, see][]{fortney.2007.prafommsiat,kennedy.2008.pfasvmslfgp,hansen.2012.mtafnpi,kretke.2012.idsstm,chiang.2013.mensfcs,swift.2013.cckikpfcpstg}, but the likelihood seems small that any simple answer for this big question could emerge. In the context of this uncertainty, we explore two representative examples. We consider period $P$ as a population-defining parameter simply because it is closely tied to what we can detect; it may not be the most fundamental physical parameter. We suspect the bolometric flux the planet receives from the host star (as parameterized by $T$) may be important for us because MEarth is most sensitive to planets larger than 2\rearth\ and that are very close-in to their stars. In the \kepler\ context, \citet{rogers.2011.fsle} and \citet{lopez.2012.temspssaksb} have argued that close-in $>2$\rearth\ planets, most of which will have gaseous outer layers, have been sculpted in large part by thermal evolution and atmospheric mass loss. The dependence of these processes on the amount of bolometric (and UV) flux that a planet receives motivates us to explore $T$ as a population-defining parameter in the following analyses.}

\subsection{MEarth found no Jupiter-sized Exoplanets}\label{s:jupiters}
We do not explicitly simulate our sensitivity to $>4$\rearth\ exoplanets. However, Figure \ref{f:sensitivity} still encodes MEarth's ability to detect gas giants. Transits of Jupiter-sized (11\rearth) exoplanets are 8\% deep for MEarth's largest targeted M dwarfs (0.35\rsun), and could be as deep as 100\% for the smallest M dwarfs. Planets this big could be detected at extremely high significance with a single MEarth observation. Our Jupiter sensitivity is probably very close to the survey window function in Figure \ref{f:sensitivity}, and is definitely bounded between the window function and the Neptune-size (4\rearth) sensitivity curve. Conservatively, we'll assume the Jupiter sensitivity lies halfway between these two curves, in log space.

Figure \ref{f:sensitivity} then indicates that if every M dwarf hosted exactly one Jupiter-sized planet in a 1.0 day orbit, MEarth should have found 70 such planets. By inverting a Poisson distribution, we can translate this statement into a 95\% confidence upper limit that  the occurrence of such 1-day Jupiters is $<0.042$ planets/star in our sample. This upper limit is not surprising. Radial velocity surveys show that about 1\% of solar-type stars host of hot Jupiters \citep{wright.2012.fjonss} and that giant planet occurrence (across all periods) decreases toward less massive stellar hosts \citep{johnson.2010.gposmp}. Only two warm Jupiters are known to transit M dwarfs \citep{johnson.2012.cckidkj, triaud.2013.wgtcd}, and for both systems the 0.6\msun\ host stars are so massive that they sit at the boundary of the K and M spectral types.

MEarth can also place an upper limit on the presence of Jupiters in orbits approaching M dwarfs' habitable zones. From the right panel of Figure \ref{f:sensitivity}, if every star hosted a Jupiter-sized planet with a zero-albedo equilibrium temperature of 300K, MEarth should have found 7. Our non-detection of any 300K Jupiters translates to $<0.44$ planets/star for such planets at 95\% confidence, again showing contradiction neither with M dwarf radial velocity surveys \citep{endl.2003.ddpsuht,johnson.2010.gposmp,bonfils.2013.hssepxms} nor with microlensing surveys \citep{sumi.2010.cnpocnc, gould.2010.fssgbslfhme2}.

Our findings from MEarth agree with other surveys that Jupiter-sized planets are uncommon around M dwarfs. We do not placer deeper limits than other surveys, although we do probe a less massive population of potential stellar hosts than they do. To be sensitive to small planets, MEarth observes a small number of stars with high photometric precision. Placing deeper constraints on the occurrence of Jupiter-sized planets can be better achieved by those surveys that monitor a larger number of M dwarfs, such as PTF/M-Dwarfs \citep{law.2012.tewbdstpam} or the WFCAM Transit Survey \citep{nefs.2012.fuembwts}.

\subsection{MEarth found no Neptune-sized Exoplanets}\label{s:neptunes}
MEarth is very sensitive to transits of planets the size of Neptune; the 4\rearth\ sensitivity curve is within a factor of two of the observational window function (the best we could possibly do) across the period and temperature ranges plotted in Figure \ref{f:sensitivity}. In light of this sensitivity, how does our lack of detections of transiting Neptunes compare with known populations?

In particular, the two Neptune-sized exoplanets that transit nearby early M dwarfs, Gl436b and GJ3470b, appear to define a family that we should address. Gl436b \citep{butler.2004.npond,gillon.2007.dtnn} is a 23\mearth, 4.2\rearth\ planet transiting a 0.45\msun, 0.46\rsun\ M dwarf,  with a 2.6 day orbital period that corresponds to a zero-albedo equilibrium temperature of 640K \citep{maness.2007.dnp, torres.2007.tehstsemlmsrpp}. GJ3470b \citep{bonfils.2012.utnd3dwhvctwtp} is a 14\mearth, 4.8\rearth\ planet transiting a 0.54\msun, 0.57\rsun\ M dwarf, with a 3.3 day orbital period that corresponds to a zero-albedo equilibrium temperature of 680K \citep{demory.2013.so3vlnpomd}. Both planets were detected initially with radial velocities and later found to transit. 

Figure \ref{f:sensitivity} indicates that MEarth should have found 20 Neptune-sized planets in 3 day periods if every one of our targets hosted such a planet, or 60 Neptune-sized planets with 650K temperatures, if every star had one. The corresponding 95\% confidence upper limits are $<$0.15 planets/star or $<$0.06 planets/star, depending on whether orbital period or the flux received from the star more genuinely define the population. For comparison, on the basis of two radial velocity planet detections from HARPS, \citet{bonfils.2013.hssepxms} report that $3^{+4}_{-1}\%$ of stars in their M dwarf sample host planets with minimum masses ($m\sin i$) between 10 and 100\mearth\ and orbital periods of 1 to 10 days. On the basis of one transit detection,  \citet{dressing.2013.orspass} find that the \kepler\ early M dwarfs host $0.004^{+0.0062}_{-0.002}$ planets per star, for planets with radii of $4.0-5.7$\rearth\ and periods $<$10 days. Determining whether the difference between these two numbers can be attributed to features in the densities for planets in this regime \citep[see][]{wolfgang.2012.epmrikhsor} will require a more thorough analysis than we wish to present here and a reduction in the Poisson uncertainties in the measurements. The upper limits from MEarth do not contradict either of these measurements.

For a consistency check, it is interesting to ask whether we would have expected the two transiting radial velocity Neptunes Gl436b and GJ3470b to have been discovered,  given the number of radial velocity planets that are not known to transit. \citet{bonfils.2013.hssepxms} compiled a list of all known planets orbiting M dwarfs. This list contains 10 non-transiting planets detected in radial velocity surveys with minimum masses in the range of $5 < m\sin i < 40$\mearth. We supplement this list with the two transiting systems and add up the values of the {\em a priori} transit probability (= $R_\star/a$) for the 12 published radial velocity planets in this mass range. We find that the total expected number of radial velocity planets that should transit is 0.5. An analysis using the planets and parameters listed in the {\tt exoplanets.org} database \citep{wright.2011.eod} yields the same result.  {The Poisson probability of having $\ge2$ detections when 0.5 are expected is only 9\%. Gl436b and GJ3470b themselves may have been slightly lucky transit detections, and the probability is low that additional M dwarf transiting planets still lurk among the published radial velocity detections.} 

\begin{figure*}[htbp]
   \centering
   \includegraphics[width=0.3\textwidth]{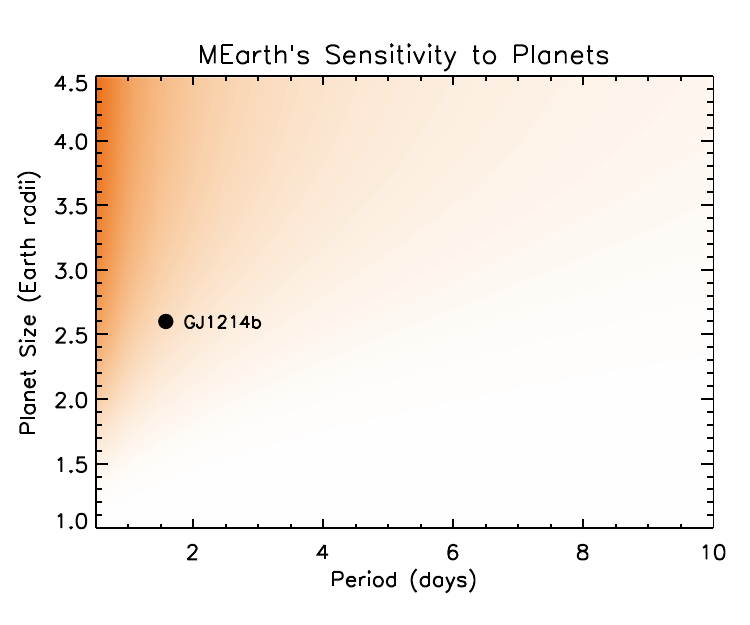} 
   \includegraphics[width=0.3\textwidth]{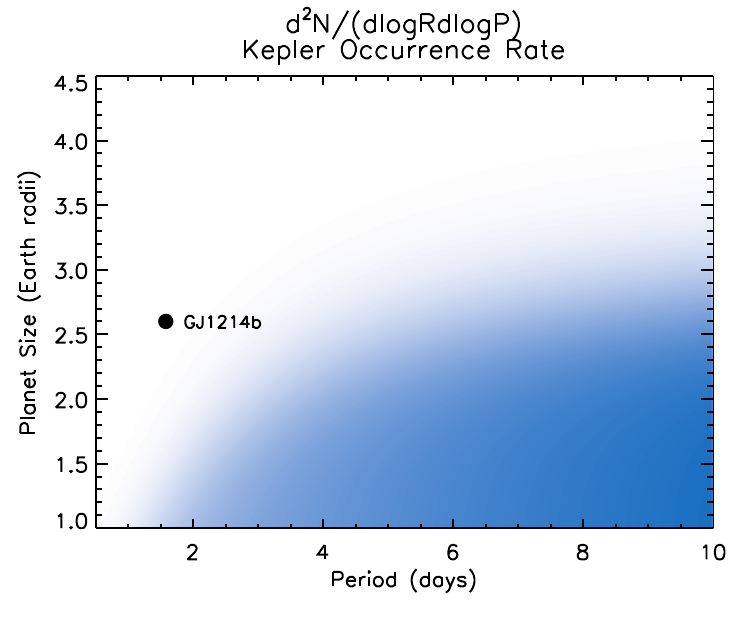} 
   \includegraphics[width=0.3\textwidth]{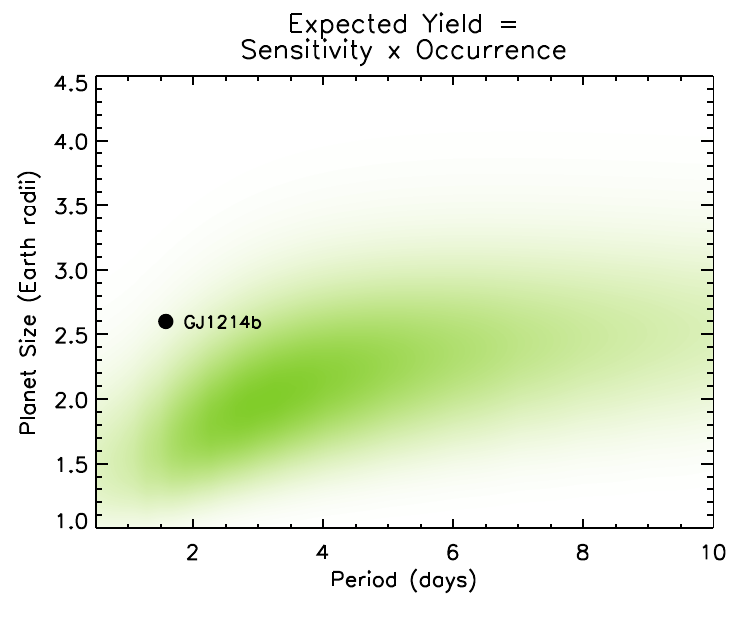} 
  
   \caption{{\em Left:} MEarth's sensitivity (in orange), as a function of planet radius and orbital period. This panel is a 2-dimensional representation of exactly the same information as in the left plot of Figure \ref{f:sensitivity}. {\em Middle:} The occurrence rate (in blue) of planets from \kepler\, for early M dwarf hosts and showing an analytic fit for $d^2N/(d\log R~d\log P)$ from the data of \citet{dressing.2013.orspass}. {\em Right:} The product of the other two panels (in green), showing the probability distribution for MEarth's expected yield (and the integrand in Eq. \ref{e:nexpected}). The integral over the parameter space shown in this panel indicates MEarth should have found $\periodexpection\pm\periodexpectionerror$ planets/year. The color scale is linear in all three panels.}
   \label{f:contours_period}
\end{figure*}

\begin{figure*}[htbp] 
   \centering
   \includegraphics[width=0.3\textwidth]{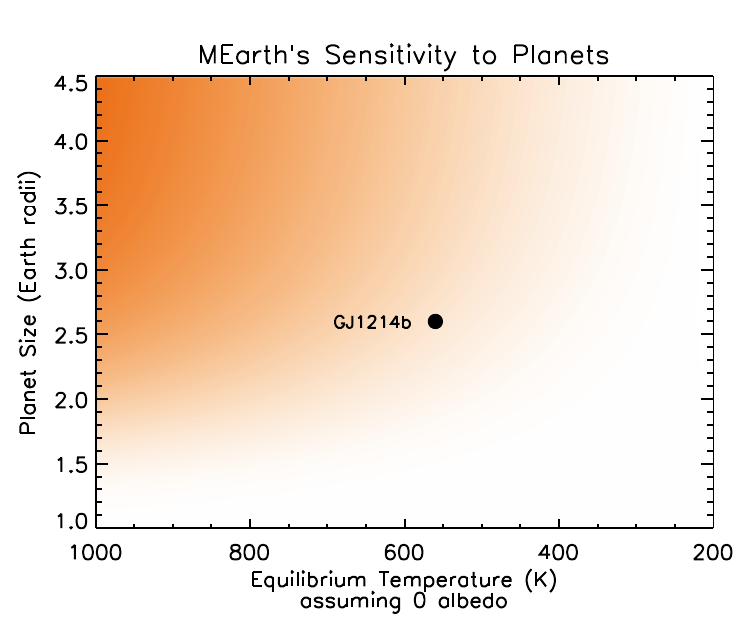} 
   \includegraphics[width=0.3\textwidth]{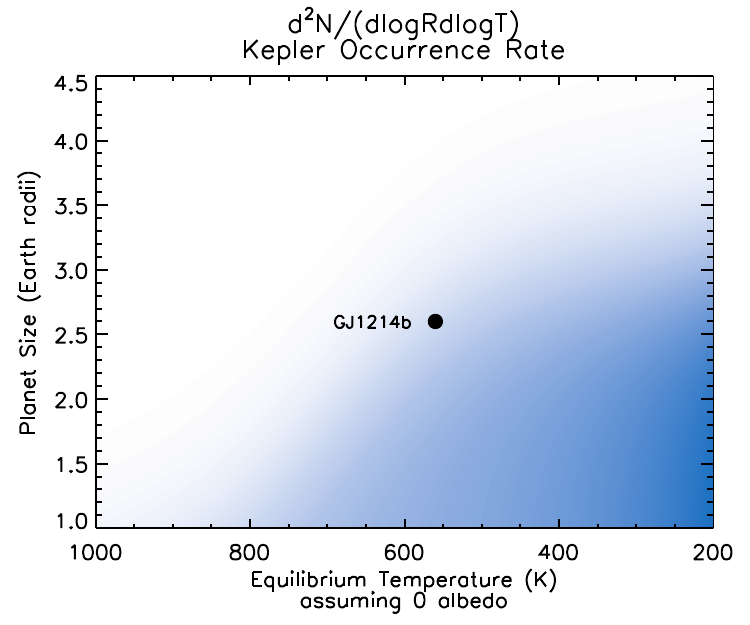} 
   \includegraphics[width=0.3\textwidth]{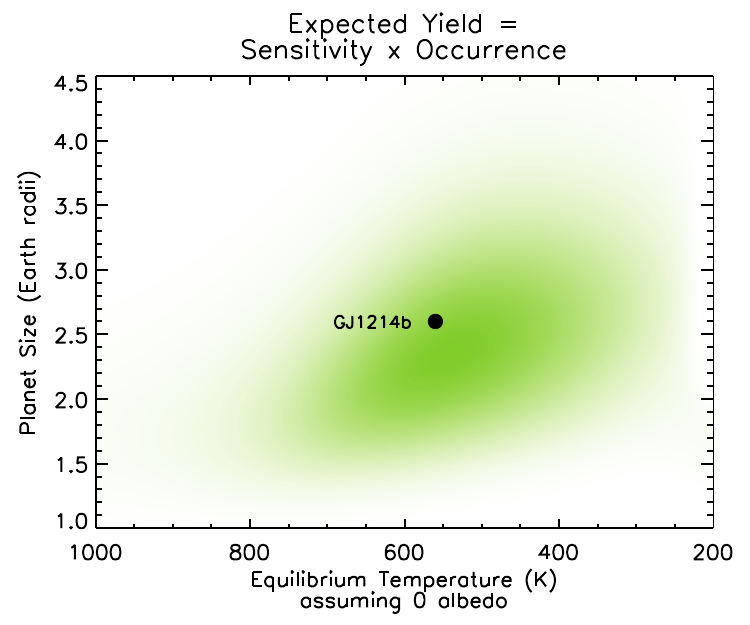} 
   \caption{The same as Figure \ref{f:contours_period}, except treating the planet occurrence distribution as a function of equilibrium temperature, $d^2N/(d\log R~d\log T)$, as fixed between \kepler's early M dwarfs and MEarth's mid-to-late M dwarfs. The integral over the parameter space shown indicates MEarth should have found  $\tempexpection\pm\tempexpectionerror$ planets/year. The color scale is linear in all three panels, but not the same as in Figure \ref{f:contours_period}.}
   
      \label{f:contours_temperature}
\end{figure*}

\subsection{MEarth found GJ1214b, a warm sub-Neptune}\label{s:gj1214b}
In \S\ref{s:jupiters} and \S\ref{s:neptunes}, we made rough statements based on visual inspection of the sensitivity curves in Figure \ref{f:sensitivity}. These point estimates effectively treat the underlying planet populations as $\delta$-functions in radius-period-temperature space. Directly integrating MEarth's sensitivity against hypothesized occurrence rate distributions is a more satisfying approach, and one that we adopt in this section to interpret our single 2.7\rearth, 1.6 day, 570K planet detection GJ1214b. We directly compare our single planet to how many planets we should have found assuming certain extrapolations from \kepler.

Throughout, we assume GJ1214b's orbital eccentricity is $e=0$. A finite eccentricity could both alter GJ1214b's size \citep{carter.2011.tlcpxsts1} and (modestly) bias its transit probability, but the effect on our statistical inferences remain small compared to the Poisson uncertainty of a single detection. 

Following \citet{howard.2012.pow0ssfk}, we define a planet occurrence distribution function $d^2N/(d\log R~d\log P)$ as the number of planets per star per {(base 10)} logarithmic interval in radius and period. With this, we can calculate the number of planets we should have expected to find as
\begin{equation}
N_{\rm expected} = \int\int \frac{ S(R,P) \times d^2N}{d\log R~d\log P} d\log R~d\log P
\label{e:nexpected}
\end{equation}
where the integrals are carried out over the radius and period ranges of interest. The expression is identical as a function of equilibrium temperature, simply exchanging $T$ for $P$.

\citet{dressing.2013.orspass} quote occurrence rates in bins of radius and period, but the bins are large compared to the scales on which MEarth's sensitivity changes substantially. To allow smooth interpolation over the parameter space and improve the accuracy of our calculation of $N_{\rm expected}$, we were motivated to fit analytic functions to the distribution. We do so in the Appendix, using the non-binning maximum likelihood formalism of \citet{youdin.2011.ecgmak} and the updated M dwarf and planetary parameters for the cool \kepler\ Objects of Interest from \citet{dressing.2013.orspass}. The results of these fits are analytic expressions for $d^2N/(d\log R~d\log P)$ and $d^2N/(d\log R~d\log T)$, along with MCMC samples that can be used to propagate the uncertainties in these distributions forward. 

Figure \ref{f:contours_period} shows the steps leading to $N_{\rm expected}$, if we assume planetary occurrence as a function of planetary radius and orbital period -- $d^2N/(d\log R~d\log P)$ --  is exactly the same (in both shape and normalization) between \kepler's M dwarfs and the MEarth sample. The left panel shows $S(R,P)$, and the middle panel shows our fit to the \kepler\ M dwarf occurrence distribution. The right panel shows the product of the first two panels (the integrand in Equation \ref{e:nexpected}), which we refer to as MEarth's expected yield. Expressed as an average over the first four years of the MEarth survey, the integral over the plotted parameter space predicts a total yield of $N_{\rm expected} = \periodexpection\pm\periodexpectionerror$ planets/year, where the error bar represents the central $1\sigma$ range of an ensemble of $N_{\rm expected}$ estimates calculated by randomly sampling both from our perturbed $S(R,P)$ curves (\S\ref{s:stellar_errors}) and from the Markov chain for the occurrence rate fit.

This overall $N_{\rm expected}$ is technically consistent with our detection of GJ1214b; the Poisson probability of finding 1 or more planets when $4\times\periodexpection=\periodexpectionfouryears$ planets are expected is \periodpoissonfindingatleast\%. However, the shape of the yield distribution also indicates what kind of planets would be the most probable for MEarth to find. In Figure \ref{f:contours_period}, GJ1214b is set apart from the bulk of the expected yield distribution; 2.7\rearth, 1.6 day planets are exceedingly rare in the \kepler\ sample. If the hypothesis that $d^2N/(d\log R~d\log P)$ stays the same between MEarth and \kepler\ holds, then the one planet that MEarth found should probably have had a longer period than GJ1214b, or been a bit smaller.

{
The agreement is much better if we assume that planet occurrence as function of radius and incident stellar flux  -- $d^2N/(d\log R~d\log T)$ -- is the same across stellar samples. Such a scaling is motivated by the dominant role thermal evolution and mass loss likely play in the histories of $2-4$\rearth\ planets \citep[][]{rogers.2011.fsle,lopez.2012.temspssaksb}. }Figure \ref{f:contours_temperature} shows the results if we focus on equilibrium temperature $T$ instead of period $P$. Because MEarth stars are on average cooler and smaller than the \kepler\ M dwarfs, this has the effect of bringing planets closer in to their host stars, making them easier for MEarth to detect. Integrating $S(R,T)$ against $d^2N/(d\log R~d\log T)$ achieves {a total $N_{\rm expected} = \tempexpection\pm\tempexpectionerror$ planets/year, in broad agreement with our single discovery in four years. The peak of the expected yield probability distribution closely matches GJ1214b's 2.7\rearth\ radius and 570K equilibrium temperature, unlike the case in Figure \ref{f:contours_period}. }

Of the two hypotheses, keeping the planet occurrence distribution fixed as a function of $T$ appears to be the better match. However, both are consistent with MEarth finding one planet in our first four years. As surveys like MEarth become more sensitive, it will be possible { to make more quantitative statements about the differences that} exist between the planet populations around mid-to-late M dwarfs and early M dwarfs. On the main sequence, the M spectral class spans a factor of 5 in stellar mass (compared to the factor of 2.5 spanned by FGK dwarfs combined). Understanding the differences across this range would provide a strong lever arm to probe the environmental dependence of planet formation. \citet{dressing.2013.orspass} may have seen hints of such differences within the \kepler\ sample itself, although their sample of very late M dwarfs is small.

\section{Discussion: Modifying MEarth}\label{s:future}
MEarth was never designed to be a statistical survey; it was designed to detect individual systems favorable for follow-up. We can use the machinery developed in this work to address how to better achieve that goal, how to increase the ``planets/year'' that MEarth could find. \kepler\ has told us planets are bountiful;  how do we reap the harvest? Inspection of  Figures \ref{f:contours_period} and \ref{f:contours_temperature} suggests that marginal improvement in MEarth's ability to detect smaller planets or cooler planets could dramatically increase the survey's yield. In this section, we describe changes to the MEarth strategy that could exploit this opportunity. Our conclusions may have broad impact for current and future surveys targeting M dwarfs, particularly those such as APACHE \citep[][]{giacobbe.2012.ptspacsfwiaps} with designs and strategies similar to MEarth's.

How do we increase sensitivity to smaller planets? Both the equipment and the observing strategy of MEarth were tailored toward detecting $>2$\rearth\ planets. To find smaller planets, we need to gather more photons during transit. Working with the existing MEarth observatory, we could achieve this goal either by exposing {longer} with each telescope during each pointing or by using {more} telescopes to monitor each star. The former is dangerous, as systematic noise sources below MEarth's current photon noise limit could potentially counteract the benefits of the extra exposure time. The latter strategy is more compelling. With the exception of a common mode effect introduced by precipitable water vapor \citep{irwin.2011.amefcsrpfmfmts,berta.2011.gsssvtsap, berta.2012.tdmsndbcsd}, many noise sources should be uncorrelated between telescopes, allowing our noise to bin down with the number of telescopes \citep[see][]{charbonneau.2009.stnls}. Increasing the number of telescopes observing each star  ($n_{\rm tel}$) decreases the total number of stars that can be observed (by a factor of $n_{\rm tel}$) but also shrinks the minimum detectable planet radius for each (by $n_{\rm tel}^{-0.25}$).  We vary $n_{\rm tel}$ and repeat the calculations in \S\ref{s:gj1214b}. We find $n_{\rm tel} =2$ optimizes the overall $N_{\rm expected}$, for which a 1.7\rearth\ planet becomes as detectable as a 2\rearth\ planet was originally. MEarth's old sensitivity so closely skirts the cliff in occurrence to make the tradeoff of observing fewer stars more deeply worthwhile. {In reality, the $n_{\rm tel}^{-0.25}$ scaling is unlikely to hold beyond 2--3 telescopes. When pushing to planets smaller than 1.7\rearth with MEarth's instrumentation, our ability to correct for the precipitable water vapor systematic would become a major concern \citep{berta.2012.tdmsndbcsd}. Scintillation would also impose a limit for bright stars and shallow depths, although scintillation may be suppressed by the extra baseline of distributing aperture over multiple telescopes \citep[see][]{young.1967.peavcrts}. For 1.7\rearth\ planets we consider here, transit are deep enough (4 millimags for a 0.25\rsun\ star) that we probably do not yet reach the multiple-telescope scintillation limit.}
 
How do we increase sensitivity to cooler, longer period planets? The probability that multiple transits will be observable from the same site within one season shrinks dramatically with increasing periods. MEarth's best strategy for longer period planets is to reliably detect planets from single transit events, by using our realtime trigger. This mode is fully functional; we have {blindly} rediscovered single transits of GJ1214b at high significance with it. We did not include simulations of the realtime trigger in the sensitivity estimate in \S\ref{s:method} because we have neither yet confirmed a new transiting planet found through the realtime trigger nor followed up on candidate triggered events sufficiently to know they were not real.

{Here, we consider in more detail what the realtime trigger means for MEarth's sensitivity.} We repeat the transit injection simulations of \S\ref{s:method} in a ``time-machine'' fashion, assessing the significance of a particular transit using only the light curve data up to and including the event. In the MISS MarPLE framework, the significance of a transit depends on how well-constrained the priors describing the star's behavior within a night are; these tighten as more nights of data are gathered \citep[see][]{berta.2012.tdmsndbcsd}. {Also, transits that occur early within a night are less significant than those that occur later, after observations have been gathered to constrain the out-of-transit baseline and the instrumental parameters for the night.} In the simulations, if a low-cadence transit exceeds a $3\sigma$ threshold we pretend the trigger would be activated, populate the remaining time left in the transit with high cadence data points, and assess the post-trigger significance. {This is a rough approximation to the actual process that occurs on-sky (a $2.5\sigma$ dip triggers the immediate start of a feedback loop to observe and continually reassess the significance of the candidate event; if the significance reaches $5\sigma$, the star is then observed continuously and unconditionally until an egress is detected or two hours elapse). For a threshold, we say that if the post-trigger significance of a single transit exceeds $5.5\sigma$ in these simulations, the planet is detected. This threshold is meant to be conservative, and it set by visual inspection of simulated events and of the histogram of single event statistics within our sample; it can be lower than the phased $7.5\sigma$ threshold because fewer independent hypotheses are being tested against the data} \citep[see][]{jenkins.2002.stecpdtp}. The simulations show that such a realtime trigger dramatically improves sensitivity to 10 day periods, especially for planets that are signal-to-noise limited in single transits (essentially everything smaller than Neptune).

For a planet detected with a single transit, we {will have essentially no knowledge of its} orbital period. One way to measure the period would be to continue to monitor the star with MEarth. This is an inefficient strategy, reintroducing the problematic bottleneck that MEarth would have to see multiple transits from the same site. External telescopes providing either continuous photometric monitoring or precise radial velocity observations {could measure the period and confirm the planet much more efficiently than MEarth could by itself.} 

\begin{figure}[htbp] 
   \centering

   \includegraphics[width=\columnwidth]{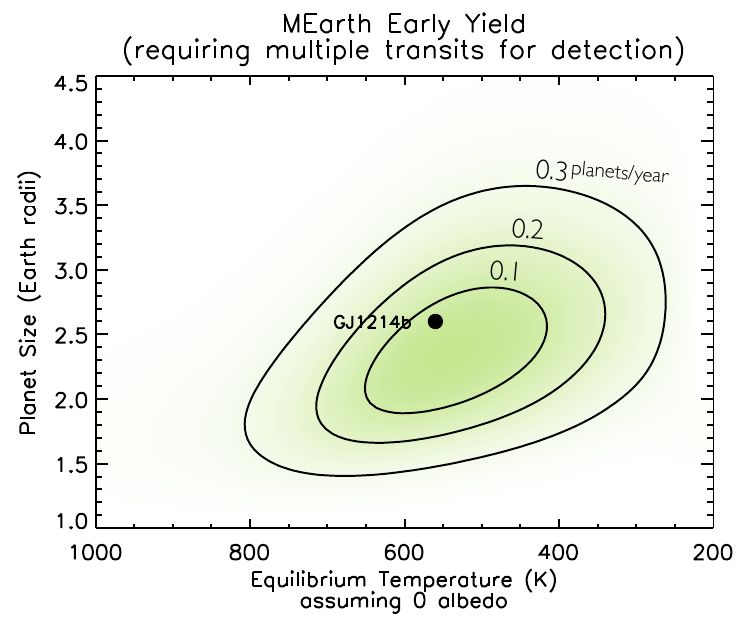} 
   \includegraphics[width=\columnwidth]{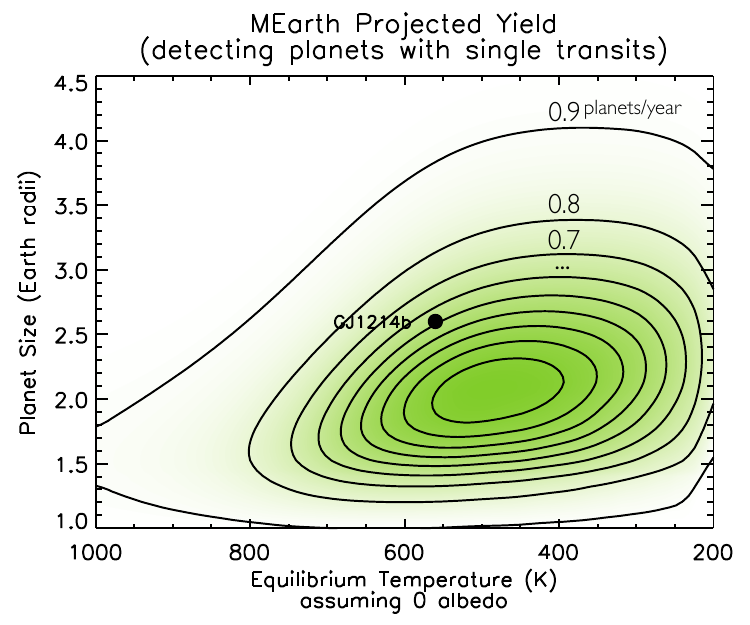} 
  
   \caption{MEarth's planet yield, as shown in Figure \ref{f:contours_temperature} but including iso-density contours for quantitative comparison between plots.  Two scenarios are shown: that achieved from phase-folded searches of the first four years of observations ({\em top}) and that which could be achieved if multiple MEarth telescopes observe the same target, planets are reliably detected with single transits, and external facilities can be used to measure their periods. Each contour is drawn such that the integral over the enclosed region corresponds to a fixed number of ``planets/year'' that MEarth should be finding. The contours increase outward in intervals of 0.1 planets/year.}
   
      \label{f:contours_comparison}
\end{figure}

The combination of all of these improvements makes MEarth more sensitive to smaller, cooler planets. Figure \ref{f:contours_comparison} compares MEarth's yield distributions for two scenarios, both assuming that $d^2N/(d\log R~d\log T)$ is fixed {between early and late M dwarfs}. The top is the original phased search yield, the same as in Figure \ref{f:contours_temperature}. The bottom is the predicted yield if MEarth allocates two telescopes per star and makes full use of the realtime trigger to detect planets with single transits, as described above. {Because fewer stars would be observed with the multiple telescope strategy, we chose those stars wisely: we explicitly biased the target list toward closer, cooler stars (those within 20 pc and smaller than $0.25$\rsun}). In both panels of Figure \ref{f:contours_comparison}, we include iso-density contours drawn so that the integral over each contour level will enclose a given number of ``planets/year.'' These contour express the overall normalization of the yield distribution, allowing for direct comparison between the panels. {In the original scenario (top panel), the integral over the entire plotted area for the original scenario is $\tempexpection\pm\tempexpectionerror$ planets/year, as it was in \S\ref{s:gj1214b}. In contrast, with the updated scenario (bottom panel), the integral over this parameter space is $\triggertempexpection\pm\triggertempexpectionerror$. Simply by revising our operational strategy, we can increase the expected yield of the survey by a factor of $2.5\times$. }

{Ultimately, the choice of whether photometry or radial velocity will provide the most efficient route to planet confirmation for these small, long period candidates will depend on the specifics of the candidate. For context, a hypothetical candidate might be a 2\rearth, 5\mearth\ planet (corresponding to a density of 3.5 g/cm$^3$) in a 12 day (roughly habitable zone) orbit around a 0.20\rsun, 0.15\msun\ star (M5 in spectral type); such a planet would create a transit 0.8\% deep and impart a stellar wobble with a semiamplitude of 5 m/s. Brute force photometry could recover additional transits and narrow the period to a family of possibilities, allowing subsequent targeted follow-up to characterize the system. The advantage of this approach is that the photometric capabilities required to reacquire 0.8\% transits are modest (comparable to MEarth's 40 cm aperture); the disadvantage is the amount of time needed, with many hours required to have a high probability of recovering long-period planet transits. The photometric option could be achieved either with a longitudinally-distributed array of telescopes such as the Las Cumbres Observatory Global Telescope (LCOGT) network \citep{brown.2013.cogtn} or with a coordinated effort across other robotic observatories.  

On the other hand, before gathering additional photometry, precise radial velocity monitoring could be used to detect the planet's orbital motion and plan subsequent targeted transit searches. Such radial velocity confirmation would be a departure from the common strategy for transiting planets of planning radial velocity observations near known times of quadrature, and it would require the detection of very small signals: for nearby mid-to-late M dwarfs, 5 m/s is comparable both to the internal errors in 15 minute exposures with the HARPS spectrograph \citep[see][]{bonfils.2013.hssepxms} and to the astrophysical jitter that can be introduced by starspots \citep[at optical wavelengths;][]{reiners.2010.dpavmswrvm,barnes.2011.edsalpdt}. The time-correlated structure of this jitter \citep[see][]{aigrain.2012.smervvsaup,boisse.2012.stfcprviss} or the presence of multiplanet systems could further complicate the extraction of the candidate planet's signal. Importantly, however, this Doppler approach would require much less time than photometry by lifting the necessity that rare transits be observed. It could potentially be achieved either with the existing HARPS, HARPS-N, or HIRES spectrographs or with the forthcoming near-IR spectrographs CARMENES \citep{quirrenbach.2012.ciso} and HZPF \citep{mahadevan.2012.hpfsfsht}, which are specifically optimized for observing mid-to-late M dwarfs like MEarth's targets.

Either way, large investments will be required to confirm MEarth's long period, single-transit planet candidates. Such major uses of telescope time are justified because the planets that MEarth can find through the above outlined strategy will likely be, for decades to come, the very best targets for atmospheric characterization studies with the James Webb Space Telescope once it launches and the giant segmented mirror telescopes once they are constructed. }

\section{Conclusions}\label{s:conclusions}
In this paper, we simulated the sensitivity of the MEarth Project to exoplanets during its first four years. We found MEarth's detection of one exoplanet to be consistent with the planet occurrence distribution measured around M dwarfs by the \kepler\ mission \citep{dressing.2013.orspass}, given reasonable assumptions about how this distribution scales between \kepler's early M dwarfs and MEarth's cooler mid-to-late M dwarfs. \kepler\ results indicate that planets are common around M dwarfs, but these planets are concentrated at orbital separations and sizes just beyond MEarth's sensitivity. Our non-detection of planets the size of Neptune or Jupiter is  consistent with literature estimates for early M dwarfs, and provides new upper limits on the occurrence rates of these planets for mid-to-late M dwarfs.

Future work will more tightly constrain how the planet occurrence distribution changes over stellar masses. As a first hypothesis, we found here that we could satisifactorilty explain our discovery of GJ1214b by assuming that planet occurrence stays fixed with respect to a planet's radius and the flux it receives from its host star, when extrapolating between \kepler's earlier M dwarfs to MEarth's later M dwarfs.

We illustrated how targeted changes to MEarth's operational program could boost its overall expected yield, by sculpting its sensitivity to match the planet distribution observed by \kepler. By sacrificing some sensitivity to large planets that we know to be rare, we can improve our sensitivity to small, cool planets that we now know to be common. Having implemented these modifications in the fall of 2012, we eagerly await the planet discoveries they may bear. The analyses we present in this paper may also help in the design of larger, future surveys to find smaller, Earth-sized planets in the habitable zones of nearby M dwarfs. 

\acknowledgements
We thank Jason Dittmann and Elisabeth Newton for detailed discussions regarding the MEarth stellar sample, and Courtney Dressing for discussing \kepler\ M planet dwarf statistics. {We are grateful to Christopher Burke and Philip Nutzman for many conversations that led to this work, and the referee for helpful comments. }The MEarth Project is supported by the David and Lucile Packard Fellowship for Science and Engineering and from the National Science Foundation (NSF) under grants AST-0807690 and AST-1109468. We are greatly indebted to the staff at the Fred Lawrence Whipple Observatory for their efforts in construction and maintenance of the MEarth facility and would like to thank Emilio Falco, Wayne Peters, Ted Groner, Karen Erdman-Myres, Grace Alegria, Rodger Harris, Bob Hutchins, Dave Martina, Dennis Jankovsky, Tom Welsh, Robert Hyne, Mike Calkins, Perry Berlind, and Gil Esquerdo for their support. This research has made use of NASA's Astrophysics Data System.


\appendix
\label{s:appendix}
\section{Estimating Analytic Planet Occurrence Distributions from Kepler's Early M Dwarfs}

Here, we estimate analytic functions that match the occurrence of planets implied by \kepler's sample of planet candidates transiting early M dwarfs. We use the maximum likelihood approach outlined in \citet{youdin.2011.ecgmak}.  We assume both that all the planet candidates listed in \citet{dressing.2013.orspass} are real planets \citep{morton.2011.fppkpc}, and that the \kepler\ pipeline is complete above the stated $7.1\sigma$ detection threshold \citep{tenenbaum.2012.dptsftqkmd}.

\begin{figure*}[htbp] 
   \centering
   \includegraphics[width=0.3\textwidth]{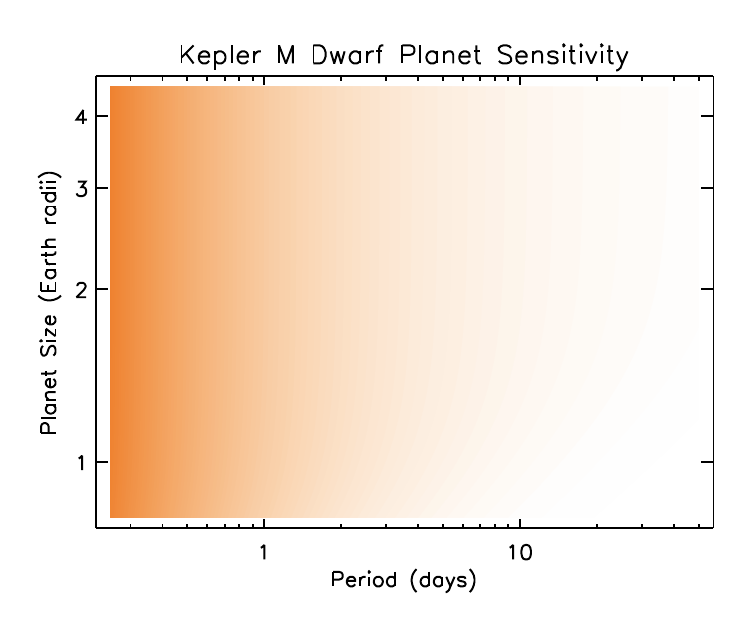} 
   \includegraphics[width=0.3\textwidth]{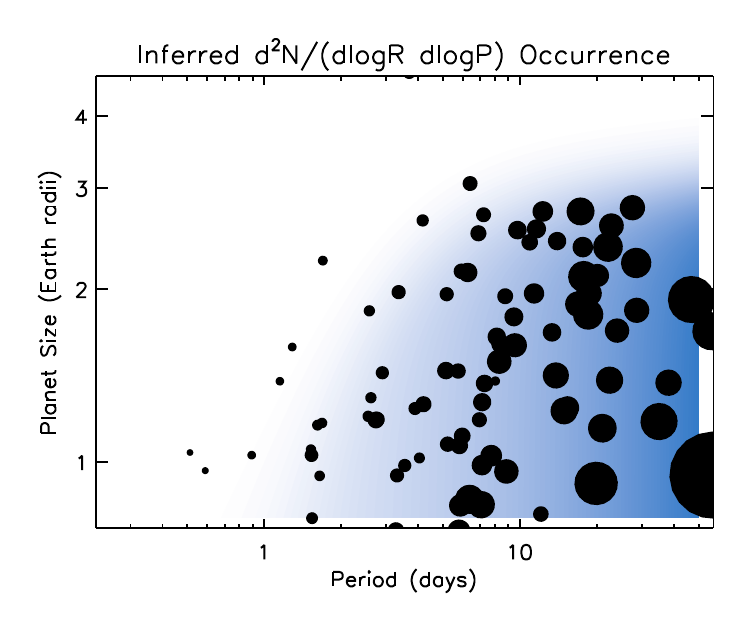} 
   \includegraphics[width=0.3\textwidth]{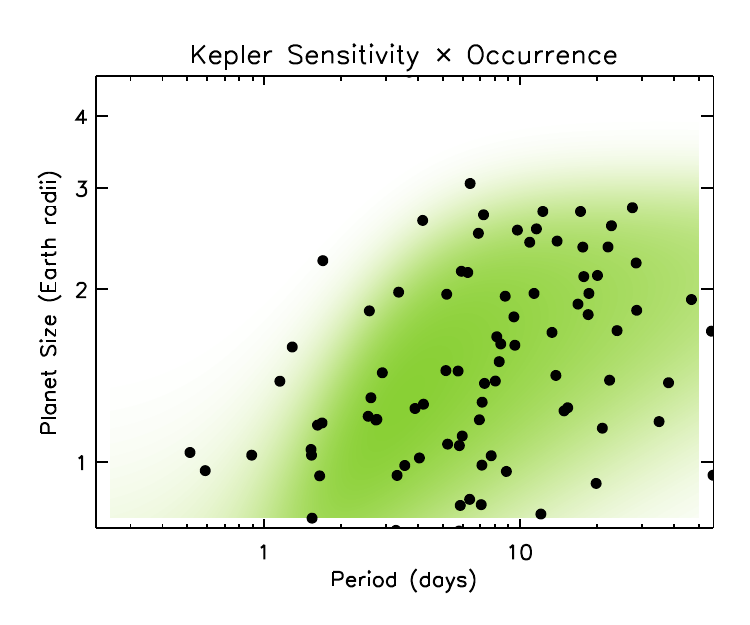} 
    \caption{Demonstration of our calculation of an analytic $d^2N/(d\log R~d\log P)$ planet distribution for \kepler's M dwarfs, using the \citet{dressing.2013.orspass} stellar/planetary parameters and sensitivity estimates. {\em Left:} Estimate of \kepler's sensitivity in orange. {\em Middle:} The inferred \kepler\ occurrence rate of planets around early M dwarfs in blue. \kepler\ planet candidates are shown as black circles whose areas scale with the number of missing planets they would represent after accounting for sensitivity effects. {\em Right:} The expected planet yield distribution from the analytic model in green; by design, this matches the distribution of the \kepler\ planet candidates (black circles).}
   \label{f:kois_period}
\end{figure*}

\begin{figure*}[htbp] 
   \centering
   \includegraphics[width=0.3\textwidth]{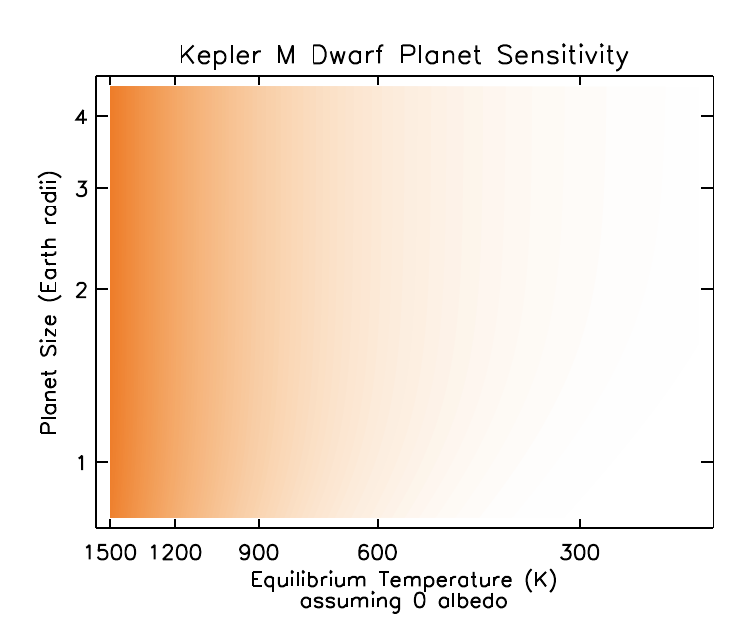} 
   \includegraphics[width=0.3\textwidth]{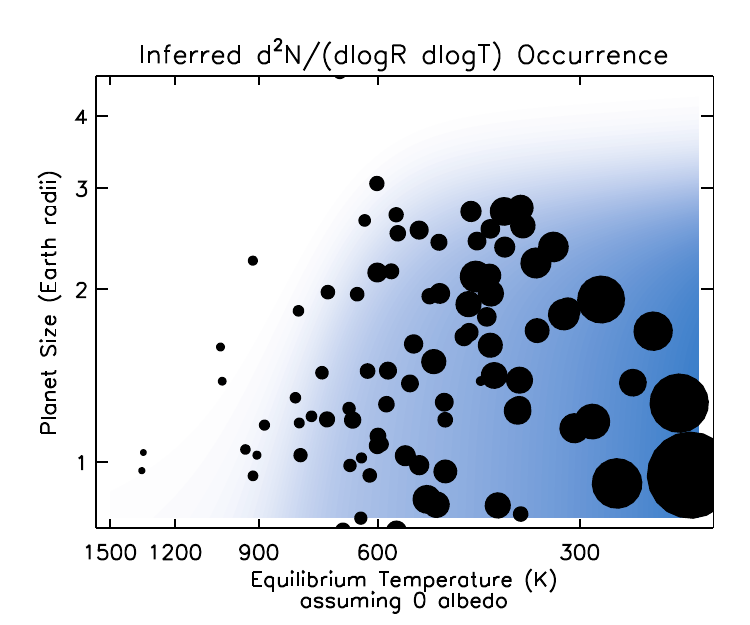} 
   \includegraphics[width=0.3\textwidth]{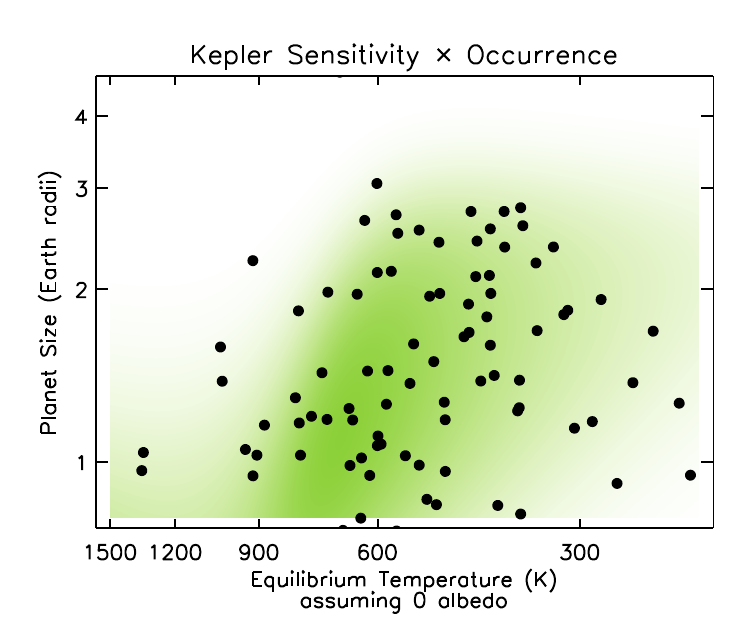} 
  
    \caption{Exactly as in Figure \ref{f:kois_period}, except for $d^2N/(d\log R~d\log T)$.}
   \label{f:kois_temperature}
\end{figure*}

We estimate \kepler's sensitivity across the parameter space by calculating representative $\eta_{\rm tra}$ and $\eta_{\rm det}$ curves from the data in \citet{dressing.2013.orspass}. We use the median stellar densities and luminosities of the M dwarf planet candidate hosts to calculate the relation between $\eta_{\rm tra}$ and $P$ and $T$. \citet{dressing.2013.orspass} estimate the fraction of their 3897 \kepler\  M dwarfs around which each planet candidate could have been detected; as in \citet{youdin.2011.ecgmak}, we fit joint power-laws for $r_{\rm det} = \eta_{\rm det}/(1-\eta_{\rm det})$ as functions of $(R,P)$ and $(R,T)$ to those per-star estimates (provided by Courtney Dressing, private communication). The resulting analytic sensitivities $S_{Kepler}(R,P)$ and $S_{Kepler}(R,T)$ are shown in the left panels of Figures \ref{f:kois_period} and \ref{f:kois_temperature}. 

In the middle panels of Figures \ref{f:kois_period} and \ref{f:kois_temperature}, we plot the \kepler\ M dwarf planet candidates as black circles. The symbol area is inversely proportional to $\eta_{\rm tra}\times\eta_{\rm det}$; with this scaling, the amount of black ink on the page should reflect the occurrence rate in that region of parameter space. We restrict our analysis to planet candidates in the plotted parameter ranges, with $0.8R_\earth < R < 4.5R_\earth$, $0.25~\mathrm{days} < P < 50~\mathrm{days}$, and $1500K > T > 200K$. The candidate KOI531.01 (4.8\rearth, 3.7 days) sits just outside this region and is not included, as we are primarily interested in planets smaller than Neptune.

We write an analytic function for the occurrence rate distribution that is purely descriptive, meant to provide a smooth approximation over the parameter space. The functional form we choose is
\begin{equation}
\frac{d^2N}{d\log R~d\log X} = \frac{(X/a)^\alpha}{1 + \exp \left[ \frac{R - R_0(X)}{\sigma} \right]} ~\mathrm{where}~R_0(X) =  \frac{b}{1 + \left(X_0/X\right)^\beta}
\label{e:df}
\end{equation}
where $X$ represents either $P$ in days or $T$ in degrees Kelvin, {and $\log$ refers to base 10 logarithms.} This expression tries to capture three qualitative features of the \kepler\ M dwarf planet candidates. The power law in the period-like parameter $(X/a)^\alpha$ captures the steady rise toward longer periods and cooler temperatures. The sigmoid factor $1/[1 + \exp (...)]$ describes a flat occurrence below a certain planet radius and steep dropoff above that radius \citep[see Figure 16 of][]{dressing.2013.orspass}. Allowing the critical radius $R_0(X)$  to increase away from the star reproduces the absence of big planets and presence of small planets at small orbital distances. 
  
{Combining this parameterized planet occurrence distribution with the above sensitivity estimate}, we write down the Poisson likelihood \kepler\ would have found the planets that it did. We maximize this likelihood using an MCMC method \citep{berta.2012.ftssgfwfchst}, and quote the mostly likely coefficients in Table \ref{t:coefs}. The right panels of Figures \ref{f:kois_period} and \ref{f:kois_temperature} show the expected yield with these coefficients. The black dots should trace this maximum likelihood yield distribution; this is the space in which the data and our model are compared. The middle panels show the inferred $d^2N/(d\log R~d\log P)$ and $d^2N/(d\log R~d\log T)$, along with the candidates weighted as described above. 

The uncertainties on the coefficients are large and strongly correlated. We do not quote these uncertainties here, but we use the MCMC samples themselves to propagate the uncertainties into the predictions in \S\ref{s:results} and \S\ref{s:future}. In Figure \ref{f:kois_temperature}, the planet temperature occurrence distribution is particularly uncertain coolward of 300K. At these temperatures, the demographics of the stars contributing to the sensitivity shifts toward the cooler ones, breaking our implicit assumption of uniform sample demographics in  $\eta_{\rm tra}$ and $\eta_{\rm det}$. Many of the planets at these large separations also have very large (of order unity) uncertainties in their estimated radii \citep[see Figure 17 of][]{dressing.2013.orspass}, which we did not account for here.

\begin{deluxetable}{lcc}
\tabletypesize{\scriptsize}
\tablecaption{The maximum likelihood coefficients for Equation \ref{e:df}, for either $X=P$ or $X=T$.\label{t:coefs}}
\tablewidth{0pt}
\tablehead{
\colhead{Coefficient} & \colhead{Period (P)} & \colhead{Temperature (T)} 
}
\startdata
 
   $a$              &          11.07 days	&	1780K	\\
   $\alpha$           &          0.51		&	-0.74 \\
   $\sigma$              &          0.29\rearth		&	0.40\rearth	\\
   $b$           &           2.9\rearth			&	 2.8\rearth	\\
   $\beta$            &           1.6		&	 -6.4\\
   $X_0$     &           2.06 days			&	744K	\\

\enddata
\tablecomments{ No uncertainty estimates are provided here; these coefficients are meant to be used for rough interpolation only. They should not be used outside the radius, period, or temperature ranges on which they were fitted.}
\end{deluxetable}

\end{document}